\documentclass[useAMS,usenatbib]{mnras}
\usepackage{times}

\usepackage{graphicx}	
\usepackage{amsmath}	
\usepackage{amssymb}	
\usepackage[usenames, dvipsnames]{color}

\newcommand {\bc}{\begin {center}}
\newcommand {\ec}{\end {center}}
\newcommand {\be}{\begin {equation}}
\newcommand {\ee}{\end {equation}}
\newcommand {\beq}{\begin {eqnarray}}
\newcommand {\eeq}{\end {eqnarray}}

\newcommand {\ergs}{{\rm erg\ \rm s^{-1}}}
\newcommand {\msun}{M_{\odot}}
\newcommand {\comment}[1]{}

\def\lbar {\lambda\hskip-5pt\raise3pt\hbox {--}}
\def\lbr {\lambda\raise2pt\hbox {\hskip-4pt{\scriptsize --}}_\C}

\renewcommand{\d}{{\rm d}}

\renewcommand{\d}{{\rm d}}

\title[Neutrino pulsars]
{Ultraluminous X-ray sources as neutrino pulsars}
\author[A. A.~Mushtukov et al.] 
{Alexander~A.~Mushtukov,$^{1,2,3}$\thanks{E-mail: al.mushtukov@gmail.com (AAM)}  
Sergey~S. Tsygankov,$^{4,2}$ 
Valery F. Suleimanov,$^{5,6}$
\newauthor
and  Juri Poutanen$^{4,7,8}$\\ 
$^1$ Anton Pannekoek Institute, University of Amsterdam, Science Park 904, 1098 XH Amsterdam, The Netherlands \\
$^2$ Space Research Institute of the Russian Academy of Sciences, Profsoyuznaya Str. 84/32, Moscow 117997, Russia \\
$^3$ Pulkovo Observatory, Russian Academy of Sciences, Saint Petersburg 196140, Russia \\
$^4$ Tuorla observatory, Department of Physics and Astronomy, University of Turku,
  V\"ais\"al\"antie 20, FI-21500 Piikki\"o, Finland \\  
$^{5}$ Institut f\"ur Astronomie und Astrophysik, Kepler Center for Astro and Particle Physics,
Universit\"at T\"ubingen, \\ Sand 1, D-72076 T\"ubingen, Germany. \\
$^{6}$ Kazan (Volga region) Federal University, Kremlevskaja str., 18, Kazan 420008, Russia.\\
$^{7}$ Nordita, KTH Royal Institute of Technology and Stockholm University, Roslagstullsbacken 23, SE-10691 Stockholm, Sweden\\
$^{8}$ Kavli Institute for Theoretical Physics, University of California, Santa Barbara, CA 93106, USA. \\
} 

\pubyear{2018}

\begin{document}
\label{firstpage}
\pagerange{\pageref{firstpage}--\pageref{lastpage}}
\maketitle

\begin{abstract}
The classical limit on the accretion luminosity of a neutron star is given by the Eddington luminosity. 
{The advanced models of accretion onto magnetized neutron stars account for the appearance of magnetically confined accretion columns and allow the accretion luminosity to be higher than the Eddington value by a factor of tens.
However, the recent discovery of pulsations from ultraluminous X-ray source (ULX) in NGC 5907 demonstrates that 
the accretion luminosity can exceed the Eddington value up to by a factor of 500. }
We propose a model explaining observational properties of ULX-1 in NGC 5907 without any ad hoc assumptions.
We show that the accretion column at extreme luminosity becomes advective. 
Enormous energy release within a small geometrical volume and advection result in very high temperatures at the bottom of accretion column, which demand to account for the energy losses due to neutrino emission which can be even more effective than the radiation energy losses.
We show that the total luminosity at the mass accretion rates above $10^{21}\,{\rm g\,s^{-1}}$ is dominated by the neutrino emission similarly to the case of core-collapse supernovae. 
We argue that the accretion rate measurements based on detected photon luminosity in case of bright ULXs powered by neutron stars can be largely underestimated due to intense neutrino emission. 
The recently discovered pulsating ULX-1 in galaxy NGC~5907 with photon luminosity of $\sim 10^{41}\,\ergs$ is expected to be even brighter in neutrinos and is thus the first known Neutrino Pulsar. 
\end{abstract}


\section{Introduction}

Ultraluminous X-ray sources (ULXs) are point-like X-ray sources of X-ray luminosity $>10^{39}\,\ergs$. ULXs are associated with the star forming regions \citep{2013MNRAS.432..506P} and likely are young objects. The phenomenon of ULXs was traditionally explained by accretion onto black holes (BHs) of stellar \citep{2007MNRAS.377.1187P} or intermediate masses \citep{1999ApJ...519...89C}. However, the recent discovery of coherent pulsations from ULXs \citep{2014Natur.514..202B,Israel17MN,2016ApJ...831L..14F,Israel17Sci} shows that some of them are accreting neutron stars (NSs) at extreme mass accretion rates.

The classical theoretical limitation for accretion luminosity  of a NS of mass $M$
is given by the Eddington value, which corresponds to the case of spherical accretion and gravity compensated by radiative force: $L_{\rm Edd}=4\pi GM m_{\rm p}c/[\sigma_{\rm T}(1+X)]\approx 1.8\times 10^{38}(M/1.4\msun)\,\ergs$,
where $m_{\rm p}$ is a mass of proton, $\sigma_{\rm T}$ is Thomson scattering cross-section and $X$ is the hydrogen mass fraction. 
The accretion luminosity of highly magnetized NSs, where surface $B$-field strength $\gtrsim10^{12}\,{\rm G}$, can exceed the Eddington value because the accreting material is confined by strong magnetic field forming accretion column \citep{BS1976} and cross-section for photon scattering by electrons is reduced in strongly magnetized plasma \citep{1979PhRvD..19.2868H,2006RPPh...69.2631H,2016PhRvD..93j5003M}. However, extremely high mass accretion rate leads to high temperatures inside the column, {when the typical photon energy becomes comparable to the cyclotron energy} and the scattering cross-section is close to the Thomson scattering cross-section \citep{2015MNRAS.454.2539M}.
As a result, luminosity of accretion column can hardly be higher than $10^{40}\,\ergs$ even in the case of strong $B$-field at the NS surface.

The accretion luminosity in particular case of ULX-1 in NGC 5907 was detected to be $~10^{41}\,\ergs$ \citep{Israel17Sci,2017ApJ...834...77F}.
Thus, the accretion luminosity of magnetized NS can exceed the Eddington value up to by a factor of 500 in pulsating ULXs. 
Explanation of such a high luminosity in combination with other observables  became an unbearable challenge for the existing theoretical models  \citep{BS1976,2015MNRAS.448L..40E,2015MNRAS.449.2144D,2015MNRAS.454.2539M,2017MNRAS.468L..59K}. 
In order to explain the data a number of unobvious assumptions like complex magnetic field structure and strong radiation beaming had to be made (see \citealt{Israel17Sci} for details).

In this paper we propose a model explaining observational properties of ULX-1 in NGC 5907 without any ad hoc assumptions.
We show that the optical thickness of accretion column across $B$-field lines at mass accretion rates above $few\times 10^{20}\,{\rm g\,s^{-1}}$ become high enough to cause strong advection. Enormous energy release within a small geometrical volume and advection result in very high temperatures at the bottom of accretion column, which demand to account for the energy losses due to neutrino emission \citep{1967ApJ...150..979B} which can be even more effective than the radiation energy losses \citep{1992PhRvD..46.4133K,1994A&AT....4..283K}.
Here we show that the total luminosity at the mass accretion rates above $10^{21}\,{\rm g\,s^{-1}}$ is dominated by the neutrino emission similarly to the case of core-collapse supernovae. 
We argue that the accretion rate measurements based on detected photon luminosity in case of bright ULX powered by NSs can be largely underestimated due to intense neutrino emission. 
The recently discovered pulsating ULX-1 in galaxy NGC~5907 \citep{Israel17Sci} with photon luminosity of $\sim 10^{41}\,\ergs$ is expected to be even brighter in neutrinos and is thus the first known Neutrino Pulsar. 

\section{The basic ideas}

At sufficiently high mass accretion rate the accretion flow onto a magnetized NS is affected by radiation pressure and is stopped at the radiation dominated shock above the NS. Below the shock the flow slowly settles down forming accretion column \citep{BS1976,LS88,2015MNRAS.447.1847M}. The matter settling inside the column is confined by the strong magnetic field leading to luminosities which may substantially
exceed the Eddington value. 

\begin{figure}
\centering 
\includegraphics[width=8.3cm]{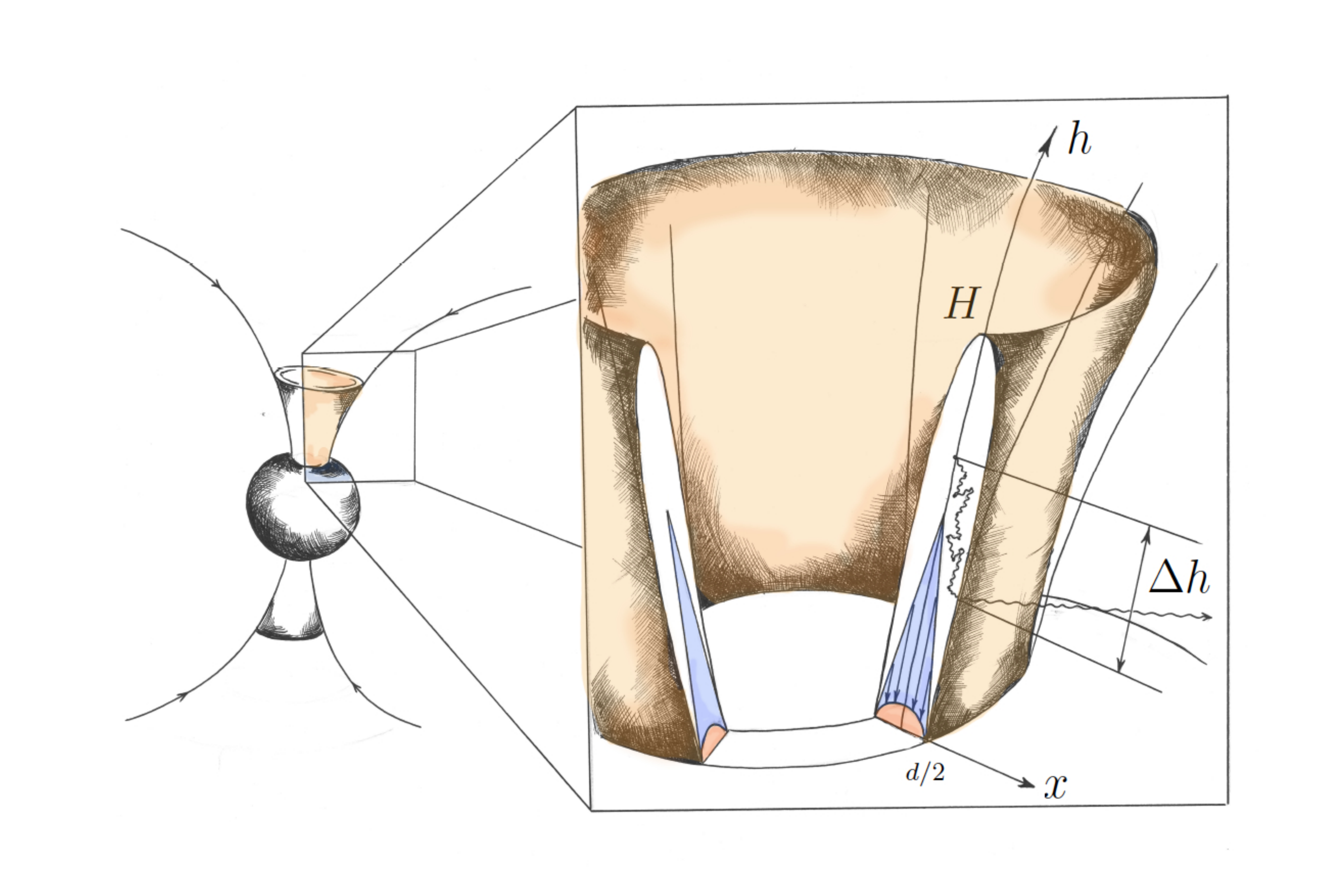}
\caption{Scheme of advective accretion column of height $H$. Advection divides accretion column into a few regions: the region producing photons responsible for the X-ray luminosity \textit{(white)}, the region which advects photons to the base of the column without their emission from accretion column walls \textit{(blue)} and 
the region producing neutrino emission out of energy advected from upper parts of the accretion column \textit{(orange)}.
}
\label{pic:scheme}
\end{figure}

The internal structure and the temperature of the accretion column are defined by the mass accretion rate and the optical thickness of accretion flow. The prime mechanism of opacity in accretion columns is Compton scattering, which is affected by strong magnetic field \citep{2016PhRvD..93j5003M}. According to the mass conservation law, the optical thickness across accretion column can be roughly estimated as
\be
\tau\approx \rho\kappa_{\rm e} d \approx \frac{\dot{m}d \kappa_{\rm e}}{v},
\ee
where $\rho$ is local mass density, $\dot{m}=\dot{M}/(2S)$ is a mass accretion rate onto unit area, $S\sim 10^{10}\,{\rm cm^2}$ is the area of accretion column base, $d\sim 10^4\,{\rm cm}$ is a geometrical thickness of accretion column \citep{2015MNRAS.454.2539M}, $\kappa_{\rm e}$ is opacity due to Compton scattering and $v=\beta c$ is a velocity of accretion flow. For extreme mass accretion rates ($\dot{M}\gtrsim 10^{20}\,{\rm g\,s^{-1}}$) the optical thickness $\tau \simeq 3\times 10^3\,\dot{m}_{10}d_4\kappa_{\rm e}/\beta \gg 1$ (we define $Q=Q_x10^x$ in cgs units if not mentioned otherwise). 
The typical time of photon diffusion from the centre of accretion column to its walls is defined by the optical and geometrical thickness of accretion channel 
$$t_{\rm diff}=\frac{\tau d}{2c}\approx 5\times 10^{-4}\,\frac{\dot{m}_{10}d^2_4\kappa_{\rm e}}{\beta}\,\,\,{\rm s}$$
and can be comparable to the settling time scale in the accretion column. 
Photons originating at one height inside the column would be emitted from the column walls at lower height due to advection, with the difference  being (see Fig.\,\ref{pic:scheme}) 
\be
\Delta h = t_{\rm diff}v = \frac{\dot{m}(d/2)^2 \kappa_{\rm e}}{c}
\approx 2.5\times 10^6\,\dot{m}_{10}d^2_4 \left(\frac{\kappa_{\rm e}}{\kappa_{\rm T}} \right)\,\,\,{\rm cm},
\ee
where $\kappa_{\rm T}\approx 0.34\,{\rm cm^2\,g^{-1}}$ is the Thomson scattering opacity in a non-magnetic case.
Because 
$$\dot{m}_{10}\approx \frac{\dot{M}_{20}}{2S_{10}}\approx \frac{1}{4}\frac{L_{40}R_6}{M_{1.4}S_{10}}$$
we get 
\be\label{eq:Delta_h}
\Delta h \approx 6\times 10^5\,\frac{L_{40}d_4 R_6}{M_{1.4}S_{10}} 
\left(\frac{\kappa_{\rm e}}{\kappa_{\rm T}} \right)\,\,\,{\rm cm},
\ee
where $L_{40}=L_{\rm tot}/10^{40}\,{\rm erg\,s^{-1}}$ is the total accretion luminosity, {$M_{1.4}=M/(1.4\,M_\odot)$ is dimensionless mass of a NS} and $\kappa_{\rm e}/\kappa_{\rm T}\lesssim 1$.
The typical $\Delta h$ can be comparable to accretion column height $H$ and even exceed it at extremely high mass accretion rates, which means that the accretion flow in the column is advection dominated and most of the photons are carried by the accretion flow down to the base of the column. It is interesting that $\Delta h$ does not depend on the velocity of the accretion flow because variations of advection velocity is fully compensated by variations of diffusion time related to the local optical depth.
{Strong advection changes the basic features of accretion column and the earlier model of accretion column based on diffusion approach \citep{2015MNRAS.454.2539M} is not valid at luminosity $\gtrsim few\times 10^{40}\,\ergs$. 
Advection prevents photons from leaving the system through the walls of the column and results in a way higher radiation energy density at its base. At sufficiently high temperatures the energy losses due to neutrino emission become more effective than the energy losses by photons which are blocked inside the column.}

{
The gravitational energy of accreting matter is transformed into the kinetic energy and thermal energy of the gas and radiation.  
In the case of extreme mass accretion rates, advection determines the local energy density and pressure $\varepsilon_{\rm acc}=\varepsilon_{\rm tot}+P_{\rm tot}$. 
Because the advection process confines photons in accretion column, 
the local energy budget $\varepsilon_{\rm acc}(h)$ at height $h$ is caused by gravitational and kinetic energy release in upper regions of the column (within the height interval $[h,h+\Delta h]$):
\be 
\varepsilon_{\rm acc}(h) \gtrsim \frac{1}{c\beta(h)} \int\limits_{h}^{h+\Delta h} {\rm d}x\,{Q^+_{\rm acc}(x)}
+\varepsilon^*_{\rm rad}(h),
\ee
where 
\beq 
Q^+_{\rm acc}(h)&=&
\frac{\d}{\d h}\left[-\frac{GM\dot{m}}{R+h}+\frac{\dot{m}v^2}{2} \right] \nonumber \\
&=&\dot{m}\left[ \frac{GM}{(R+h)^2}+\beta c^2 \frac{{\rm d}\beta}{{\rm d}h} \right]
\eeq
is accretion heating rate {defined by variations of the gravitational potential energy and the kinetic energy} and $\varepsilon^*_{\rm rad}$ is the radiation energy density advected from the radiation dominated shock at the top of accretion column. 
Note that the radiation shock region can be responsible for significant fraction of total accretion luminosity: $L^*_{\rm sh}\approx GM\dot{M}/(R+H)$. 
Both advected energy density and pressure contribute eventually to the luminosity of the X-ray pulsar.
}


The exact distribution of radiation energy density is defined by material velocity in the accretion column, but in any case there is a region at the bottom of the accretion column, where the temperature exceeds $10^{10}\,{\rm K}$ (see Section \ref{sec:Possibility}).
\footnote{
{The extreme internal temperature above $10^{10}\,{\rm K}$ at mass accretion rates $\dot{M}\gtrsim 10^{21}\,{\rm g\,s^{-1}}$ can be obtained independently from the effective temperature at the accretion column walls and expected optical thickness across the magnetic field lines.}
} 
Such a high temperature results in energy losses by neutrino emission \citep{1967ApJ...150..979B,1994A&AT....4..283K}.

The main mechanism of the neutrino production for the mass density of $\rho<10^{7}\,{\rm g\,cm^{-3}}$ is annihilation of electron-positron pairs with appearance of neutrino--anti-neutrino pairs: $e^{-}+e^{+}\longrightarrow \nu+\overline{\nu}$ \citep{1967ApJ...150..979B}.
A strong magnetic field influences the processes of neutrino production \citep{1994A&AT....4..283K,2008JETP..107..533R} and also gives rise to synchrotron neutrino production: $e+B \longrightarrow e+\nu+\overline{\nu}$. 
At high temperature regime the pair annihilation remains the dominant process for neutrino production with the corresponding  energy losses  of
\be\label{eq:Qnu}
Q^{-}_\nu\approx 4\times 10^{24} \left(\frac{T}{10^{10}\,{\rm K}}\right)^{9}\,\,{\rm erg\,cm^{-3}\,s^{-1}}
\ee
for the case of nondegenerate plasma at temperature $>10^9\,{\rm K}$ (see Appendix \ref{App:Nu}).
The magnetic field at the bottom of accretion column has to be strong enough to confine plasma of high temperature, $B>2.5\times 10^{13}\,T_{10}^2\,{\rm G}$.  This limitation becomes even stronger if we account for the gas pressure, which can be as high as the radiation pressure at the base of the column. Indeed, the gradient of radiation pressure cannot support the accretion column because of energy losses due to neutrino emission. 

As a result, strong advection nascent at extremely high mass accretion rate results in appearing of a zone in the central regions of the accretion column, where the released energy is advected to the bottom of the column and emitted by neutrinos. 
\footnote{
{Note, that high concentration of electron-positron pairs makes the opacity due to Compton scattering even higher, which reinforces advection in accretion column. Thus, equation (\ref{eq:Delta_h}) gives a lower limit of the advection length and on the neutrino luminosity. }
}
At the same time, the outer parts of accretion column lose energy due to emission of X-ray photons responsible for the observed luminosity. 
Hence, the total accretion luminosity is given by a sum of the photon and neutrino luminosities:
\be 
L_{\rm tot}=\frac{GM\dot{M}}{R}=L_{\rm ph}+L_{\rm\nu}.
\ee 
It is defined by the total energy release within the accretion column:
\be 
L_{\rm tot}=2 ld\,\int\limits_{0}^{H}\d h\, Q^+_{\rm acc}(h)+ L^*_{\rm sh},
\ee 
while the neutrino luminosity can be estimated as a fraction of photons trapped in the advection-dominated part of the column
\be
L_{\nu}\approx 4 l\,\int\limits_{0}^{d/2}\d x\,\int\limits_{0}^{\Delta h(x)}\d h\, Q^+_{\rm acc}(h)
+f L^*_{\rm sh}, 
\ee
where $\Delta h(x)=\dot{m}x^2\kappa_{\rm e}/c$ is a height below which the photons cannot escape from the column because of advection ($\Delta h$ determines the geometry of a "blue" zone in Fig.\,\ref{pic:scheme}) and $f=\max\{0;\,1-[cH/(\dot{m}\kappa_{\rm e})]^{1/2}\}$ (determined by $\Delta h$) is a fraction of luminosity advected from the shock region to the region, where neutrino emission dominates.

At relatively low mass accretion rates ($< 10^{20}\,{\rm g\,s^{-1}}$), photons can escape from the accretion column, the neutrino luminosity is negligible and the total luminosity is dominated by photons. 
At high mass accretion rates ($>10^{21}\,{\rm g\,s^{-1}}$), there is a large region where photons are trapped and the energy is transported to the region cooled by neutrinos, resulting in the luminosity exceeding the photon luminosity (see Fig.\,\ref{pic:res}, \textit{left}).
A high neutrino luminosity implies that the mass accretion rate in bright ULXs powered by accretion onto magnetized NSs can be significantly underestimated. Moreover, the photon luminosity may have a natural upper limit of about $10^{41}\,\ergs$ because at higher $\dot{M}$ most of the energy is converted into neutrinos and an increase of photon luminosity is much slower.

\subsection{Possibility of extreme temperature at the bottom of advection dominated accretion column}
\label{sec:Possibility}

{
Let us estimate the physical conditions at the bottom of advection dominated accretion column. The gravitational energy of accreting matter transforms into kinetic energy, internal energy of the gas and radiation and rest mass energy of electron-positron pairs. 
Thus, the energy conservation law in steady flow, for which time derivatives are put equal to zero, can be represented as follows:
\be\label{eq:EnergyCons01}
\frac{\partial}{\partial h}\left[ 
\left(-\frac{\rho GM}{R+h}+\frac{\rho v^2}{2}+\varepsilon_{\rm tot}+P_{\rm tot}  + 2n_{+}m_{\rm e}c^2 \right) v
\right] = Q^{-},
\ee
where $Q^{-}=Q_{\rm ph}^{-}+Q_\nu^{-}$ is the energy lost rate due to emission of photons and neutrinos,
$$\varepsilon_{\rm tot}=\varepsilon_{\rm gas}+\varepsilon_{\rm rad}=\frac{3}{2}nk_{\rm B}T+aT^4$$
is  the total energy density, 
$$P_{\rm tot}=P_{\rm gas}+P_{\rm rad}= nk_{\rm B}T + \frac{1}{3}aT^4$$
is the total pressure, $n=(n_{\rm p}+n_{-}+n_{+})$ is number density of massive particles (barions, electrons and positrons), $n_{-}$ and $n_{+}$ are the number densities of electrons and positrons respectively.
}

{
The number densities of electrons and positrons $n_\mp$ in the equilibrium at given temperature $T$ can be found from their distributions functions \citep{1992PhRvD..46.4133K,2006RPPh...69.2631H}:
\be
n_{\mp}=\left(\frac{m_{\rm e}c^2}{\hbar}\right)^3 \frac{b}{4\pi^2}\sum\limits_{n=0}^{\infty} g_{n}\int\limits_{-\infty}^{\infty}\d p_{\rm z}f_{\mp, n}(p_{\rm z}),
\ee
where 
$$f_{\mp, n}=\left(\exp\left[\frac{E_{n}(p_{\rm z})\mp \mu_{\rm e}}{t}\right]+1\right)^{-1}$$
describe distribution of electrons and positrons over longitudinal momentum $p_{\rm z}$ at $n$-th Landau level,
$E_{n}(p_{\rm z})=(1+p_{\rm z}^2+2bn)^{1/2}$ is the electron energy,
$g_n=2-\delta_{n}^0$ is the spin degeneracy of the Landau levels,  $\mu_{\rm e}$ the chemical potential,
$t=T/(5.93\times 10^9\,{\rm K})$.
The chemical potential $\mu_{\rm e}$ can be found out from the condition of charge neutrality: $n_{-}-n_{+}=\sum_i Z_i n_i$, where $n_i$ and $Z_i$ are number densities and charges of atomic nuclei. We calculate $n_\mp$ numerically under the assumption of completely ionized hydrogen plasma.
}

\begin{figure}
\centering 
{\includegraphics[width=8.5cm]{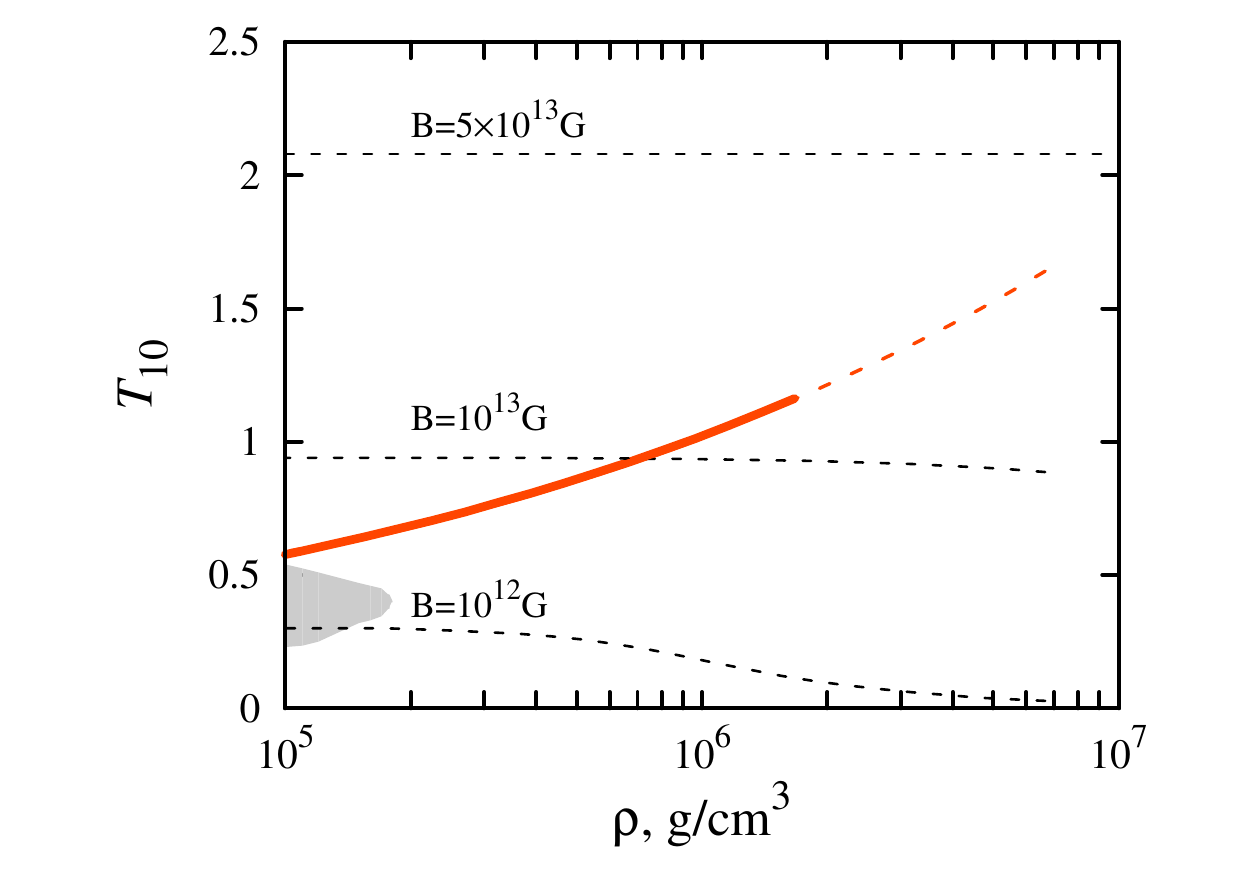}}
\caption{
{The dependence of temperature in advection dominated accretion column on barion mass density (red solid line) calculated under assumption of negligibly small energy losses due to photon emission. The black dashed lines give upper temperature limits for the regions, where accreting plasma can be confined by strong magnetic field of NS. One can see that the temperatures above $10^{10}\,{\rm K}$ can not be reached if $B\lesssim 10^{13}\,{\rm G}$.  The grey zone corresponds to the region, where the radiation pressure dominates over the gas pressure. }
}
\label{pic:res_}
\end{figure}

{
If the temperature $T\lesssim 10^{10}\,{\rm K}$, the energy losses due to neutrino emission are small and $Q^{-}\approx Q^{-}_{\rm ph}$.
Let us estimate the influence of energy losses due to photon emission $Q_{\rm ph}^{-}$. They are determined by radiation energy density $\varepsilon_{\rm rad}$ and optical thickness of accretion column $\tau$:
\be
Q_{\rm ph}^{-}\approx \frac{2c}{3d}\frac{\varepsilon _{\rm rad}}{\tau} 
\approx 2\times 10^{3}\frac{\varepsilon_{\rm rad}\beta}{\dot{m}_{10}d_4^2\kappa_{\rm e}}\,\,{\rm erg\,cm^{-3}\,s^{-1}}.
\ee
Because the radiation energy density can be estimated from above:
\be
\varepsilon_{\rm rad}< \frac{1}{v}\frac{\dot{m}GM}{R+h}\approx 6.2\times 10^{19}\frac{\dot{m}_{10}M_{1.4}}{\beta (R+h)_6}\,\,{\rm erg\,cm^{-3}}
\ee
one can put an upper limit on the photon energy losses  
\be
Q_{\rm ph}^{-} < 1.3\times 10^{23}\frac{M_{1.4}}{d_4^2 \kappa_{\rm e}(R+h)_6}\,\,{\rm erg\,cm^{-3}\,s^{-1}}.
\ee
This estimation will be even stronger if one would take into account the increase of optical thickness of accretion column due to the pair creation at high-temperature regime. The ratio of energy losses due to emission of photons to the gravitation energy release $\dot{m}GM/(R+h)^2$: 
\be
\frac{Q_{\rm ph}^{-}}{\dot{m}GM(R+h)^{-2}}< 6.8\times 10^{-2}\frac{(R+h)_6}{\dot{m}_{10}d_4^2 \kappa_{\rm e}}.
\ee
It means that only a small fraction (less than a few per cents in the case of $\dot{M}\gtrsim 10^{21}\,{\rm g\,s^{-1}}$) of energy would be lost due to photon emission from accretion column walls.
Thus, we can neglect $Q^{-}$ in equation (\ref{eq:EnergyCons01}) in the case of extreme mass accretion rates, i.e. strong advection in the column results in roughly adiabatic regime of accretion.
Because the number density of massive particles is $n=n_{\rm p}/\mu = \dot{m}/(m_{\rm p}v\mu)$, where $\mu\approx n_{\rm p}/(n_{\rm p}+n_{-}+n_{+})$ is the mean molecular weight, one can rewrite (\ref{eq:EnergyCons01}) as 
\be
\frac{\partial}{\partial h}\left[ 
-\frac{\dot{m} GM}{R+h}+\frac{\dot{m} v^2}{2}+\frac{5}{2}\frac{\dot{m}k_{\rm B}T}{\mu m_{\rm p}} + \frac{4}{3}vaT^4  + 2n_{+}m_{\rm e}c^2 v
\right] = 0,
\ee
The expression in square brackets does not depend on height $h$ and turns to zero at the infinity. Therefore
\be  
\frac{\dot{m} v^2}{2}+\frac{5}{2}\frac{\dot{m}k_{\rm B}T}{\mu m_{\rm p}} + \frac{4}{3}vaT^4  + 2n_{+}m_{\rm e}c^2 v-\frac{\dot{m} GM}{R+h}=0.
\ee
Dividing the expression by the local velocity and local number density of barions we get the equation of energy conservation per one barion:
\be 
\frac{m_{\rm p}v^2}{2}+\frac{5}{2}\frac{k_{\rm B} T}{\mu} +\frac{4}{3}\frac{v m_{\rm p}}{\dot{m}}aT^4
+\left(\frac{2n_{+}}{n_{-}-n_{+}}\right)m_{\rm e}c^2 -\frac{GMm_{\rm p}}{R+h} = 0
\ee
These equation can be rewritten in appropriate units as:
\beq\label{eq:ro_T10}
7.5\beta^2 +1.5 \frac{T_{10}^4}{\rho_7}+3.45\times 10^{-2}\frac{T_{10}}{\mu} && \nonumber \\
+8.2\times 10^{-3}\frac{2n_{+}}{n_{-}-n_{+}}
-3.1\frac{M_{1.4}}{(R_6+h_6)} &= & 0 .
\eeq
Because the number density of electrons and positrons are determined by temperature and density, equation (\ref{eq:ro_T10}) gives the  approximate relation between local barion mass density $\rho$ and temperature. Using this expression one can estimate the conditions, when the temperature becomes as high as $10^{10}\,{\rm K}$ under assumption that the photon energy losses are negligibly small (see Fig.\,\ref{pic:res_}). One can see that the accreting matter achieves temperature of $10^{10}\,{\rm K}$ at barion mass density $\sim 10^{6}\,{\rm g\,cm^{-3}}$ in the case of advection dominated accretion column.
}

{
The hot regions of accretion column should be supported against spreading over the NS surface by strong magnetic field. The critical pressure for this kind of spreading was estimated by \cite{1997ApJ...477..897B}:
\be
P_{\rm crit}\approx 3\times 10^{23} S_{10}^{1/2}B_{12}^2 T_{10}^{-1}\,\,{\rm erg\,cm^{-3}}. 
\ee
The plasma is confined by strong magnetic field if $P<P_{\rm crit}$. The corresponding upper limits of temperature for various surface $B$-field strength are given by black dotted lines in Fig.\,\ref{pic:res_}.
}

{
The gas pressure in extremely hot part of accretion column is comparable to or even higher than the radiation pressure. It happens because of high number density of electron-positron pairs.
\footnote{These conditions are similar to those in early Universe of age $10^{-4}\,{\rm s}<t<10^{-2}\,{\rm s}$ (see e.g. \citealt{1975seu..book.....Z}).}
}

\section{ULX-1 NGC~5907 as neutrino pulsar}

\begin{figure*}
\centering 
{\includegraphics[width=18.cm]{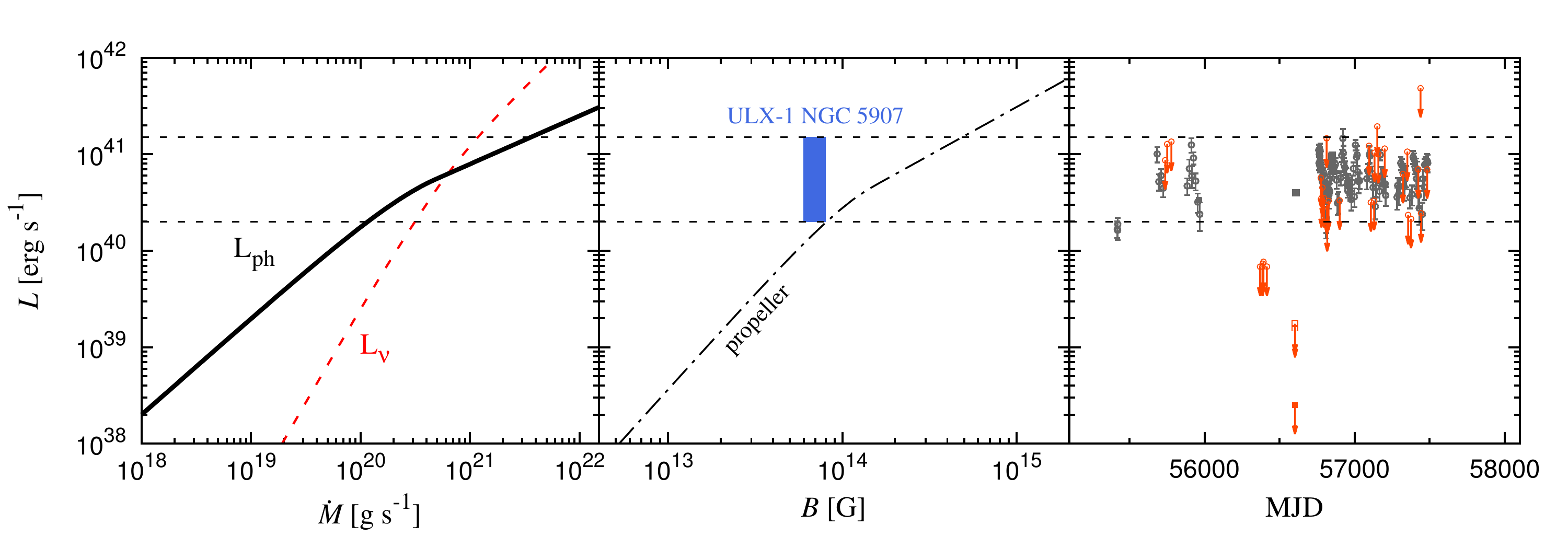}}
\caption{ \textit{Left:} The photon (black solid line) and neutrino (red dashed line) luminosities as functions of the mass accretion rate. At relatively low mass accretion rate, most of energy is converted into photons, but at very high accretion rates the neutrino luminosity  exceeds the photon luminosity. 
Parameters used in the calculations: $H=10^6\,{\rm cm}$, $d=10^4\,{\rm cm}$, $\kappa_{\rm e}/\kappa_{\rm T}=0.2$.
\textit{Centre:} The lower limit on the X-ray luminosity (neutrino emission taken into account) from the propeller effect (grey dashed-dotted line) for a pulsar with the period $P=1.137$~s. 
The likely position of ULX-1 NGC~5907 corresponding to the propeller luminosity of $2\times 10^{40}\,\ergs$ is marked by a blue vertical bar with  the width corresponding to the error on B and the height corresponding to the range of observed photon luminosities.
\textit{Right:} The light curve of ULX-1 NGC 5907 \citep{Israel17Sci} obtained by \textit{Swift} (circles), \textit{NuSTAR} (empty squares) and \textit{XMM-Newton} (filled squares). Red arrows show $3\sigma$ upper limits. The horizontal dashed lines in all three panels show the photon luminosity range, where the ULX-1 NGC 5907 has been detected.
}
\label{pic:res}
\end{figure*}

Photon advection in the accretion column and a high neutrino luminosity can explain a puzzle around a recently discovered very bright pulsating ULX in the galaxy NGC~5907, whose photon luminosity reaches a record value of $L_{\rm ph}\approx 2\times 10^{41}\,\ergs$ (see Fig.\,\ref{pic:res}, \textit{right}) \citealt{Israel17Sci}. Strong non-dipole magnetic field ($B\sim 3\times 10^{14}\,{\rm G}$ at the surface with 
the dipole component being 10 times weaker) and a strong beaming by a factor of 10 of the X-ray radiation have been proposed to explain the observed properties of this source \citep{Israel17Sci}.
The reduced dipole component of the magnetic field was required to avoid the onset of the propeller regime \citep{1975A&A....39..185I}.  
Strong beaming of the emission was required to reduce the actual mass accretion rate onto the NS to exclude super-critical disc accretion \citep{1982SvA....26...54L} and unstable accretion column \citep{2015MNRAS.454.2539M}. 
However, all these ad hoc assumptions are not needed if one would take into account physical conditions appearing at such high mass accretion rates. 
Particularly, the accretion disc becomes advective, its geometrical thickness $z$ is strongly reduced: $z\lesssim (0.3 - 0.6)R_{\rm m}$ for $\dot{M}\lesssim 10^{21}\,{\rm g\,s^{-1}}$ \citep{1998MNRAS.297..739B,2007MNRAS.377.1187P}. 
As a result, the accretion disc is not super-critical at given accretion luminosities \citep{2017arXiv1703.07005C}. 
At the same time, model introducing the maximal accretion luminosity of the columns \citep{2015MNRAS.454.2539M} is also inapplicable for the luminosities above $\sim 10^{40}\,\ergs$ because it was constructed  assuming a non-advective accretion flow in the NS vicinity. At higher mass accretion rates, the column becomes advection dominated, as we have shown above, and a significant part of the released energy is advected to the bottom of the accretion column and converted to the neutrinos. Thus, there are no reasons to assume strong beaming of the X-ray radiation, which also contradict the observed high pulsed fraction of $\sim 20\%$ \citep{Israel17Sci}. In addition the strong beaming can be excluded because of appearance of optically thick envelope at the magnetosphere of NS at extreme mass accretion rate \citep{2017MNRAS.467.1202M}. The envelope effectively smooths out pulse profiles and can result in strong thermal component in the energy spectrum, which has been detected in ULX-1 NGC 5907 \citep{2017ApJ...834...77F}.

Weak dipole component of the NS magnetic field and accretion luminosity $\lesssim 2\times 10^{40}\,\ergs$ proposed earlier \citep{Israel17Sci} can be principally excluded because of high  spin period derivative.
Variations of NS spin period are caused by accelerating and decelerating torques \citep{2016ApJ...822...33P}. 
Thus, a large period derivative puts a strong lower limit to the total accretion luminosity \citep{2017MNRAS.468L..59K}:
\beq\label{eq:Ltot_dotP}
L_{\rm tot} & > & 10^{40}\,\left(\frac{|\dot{P}|}{3.7\times 10^{-9}\,{\rm s\,s^{-1}}}\right)^{7/6}\left(\frac{B}{10^{14}\,{\rm G}}\right)^{-1/3}
P^{-7/3} \nonumber \\ && \times \frac{M}{M_\odot} R_6^{-2}I_{45}\,\,\ergs.
\eeq
However, the spin period derivative is quite uncertain in the case of ULX-1 NGC 5907.
The secular period derivative $\dot{P}_{\rm sec}\approx - 8\times 10^{-10}\,{\rm s\,s^{-1}}$ obtained from the spin period difference in observations performed in February 2003 and in July 2014 is about $10$ times smaller than the spin period derivative measured in single observations. The secular spin period derivative is affected by accretion history of the source during these 11 years (note that within this period the source was detected to be in propeller state several times, and, therefore, was spun down) and, thus, is naturally small. On the other hand, the period derivatives measured in single observations can be affected by the orbital motion in binary system.
The uncertainty in actual spin period derivative in ULX-1 NGC 5907 poses a question for future observations.

For the peak observed luminosity value, our model predicts the neutrino luminosity to exceed  the photon luminosity by a factor of 5 (see Fig.\,\ref{pic:res}, \textit{left}). 
At such $\dot{M}$ the variability of photon luminosity by a factor of three  corresponds to actual mass accretion rate variation by an order of magnitude. 
It results in a natural upper limit of the observed luminosity from such kind of sources.

If one interprets the dramatic variability of ULX NGC~5907 below $L_{\rm ph}\sim 2\times 10^{40}\,\ergs$ (see Fig.\,\ref{pic:res}, \textit{right}) as a ``propeller" effect  \citep{1975A&A....39..185I} (a similar behaviour has been detected in bright X-ray pulsars \citep{2016A&A...593A..16T} and ULX M82 X-2 \citep{2016MNRAS.457.1101T}),  then the surface magnetic field strength of the dipole component can be estimated as $\sim 10^{14}\,{\rm G}$ (see Fig.\,\ref{pic:res}, \textit{center}). It is enough to confine high temperature plasma producing neutrino emission. This value is also in agreement with restrictions given by large $\dot{P}$ and does not require strongly non-dipole configuration of the magnetic field. 

\section{Summary and discussion}

We show that accretion columns above highly magnetized NSs become advection dominated at mass accretion rates $> few\times 10^{20}\,{\rm g\,s^{-1}}$. It requires a significant revision of accretion column models \citep{BS1976,2015MNRAS.454.2539M} for the case of extremely bright ULX pulsars.
Extreme mass accretion rate and advection result in very high temperatures ($\sim10^{10}\,{\rm K}$) at the bottom of the column, which demand to account for the energy losses due to neutrino emission which becomes more effective than the radiation energy losses.

We conclude, that the recently discovered bright ULX-1 in galaxy NGC~5907 is expected to be even brighter in neutrinos and, therefore, can be considered as the first Neutrino Pulsar with the surface magnetic field strength $\sim 10^{14}\,{\rm G}$. The energy of produced neutrinos is expected to be within a range $10^5 \div 10^{6}\,{\rm eV}$ \citep{2006PhRvD..74d3006M}, which is explored by underground neutrino detectors. However, the expected neutrino energy flux from the ULX in NGC~5907 at Earth of a few times $10^{-11}\,{\rm erg\,cm^{-2}\,s^{-1}}$ corresponds to $\sim 10^{-5}$ neutrino cm$^{-2}$\,s$^{-1}$, which is below current sensitivity level of neutrino detectors. 
The most important consequence of a high neutrino luminosity is that the mass accretion rate in bright ULXs can be largely underestimated if it is based on the observed photon luminosity. It is expected that even small variability of photon luminosity can be associated with large variability of $\dot{P}$, which is defined by actual mass accretion rate.

Strong energy losses due to neutrino production limits the photon luminosity of accreting NSs from above, which would influence the luminosity function of high mass X-ray binaries, which shows a sharp cut-off above $few\times 10^{40}\,\ergs$ \citep{2012MNRAS.419.2095M}. It might be an additional evidence of NS domination in ULXs.

\section*{Acknowledgements}

AAM was supported by the Netherlands Organization for Scientific Research (NWO).
SST acknowledges the support of the Academy of Finland grant 309228. 
VFS was supported by the German Research Foundation (DFG) grant WE 1312/48-1 and the Magnus Ehrnrooth foundation.
JP was partly supported by the Foundations' Professor Pool, the Finnish Cultural Foundation, and the National Science Foundation grant PHY-1125915.
AAM, SST and VFS acknowledge support by the Russian Science Foundation grant 14-12-01287 in part of critical analysis of the observational data and fittings the data by proposed model (Section 3).
We also acknowledge the support from the COST Action MP1304.
We are grateful to Michiel van der Klis, Paolo Esposito and Dmitry Rumyantsev for a number of useful comments.


\begin{thebibliography}{}

\bibitem[\protect\citeauthoryear{{Bachetti}}{{Bachetti et al.}}{2014}]{2014Natur.514..202B}
{Bachetti} M. et al.,  2014, Nature, 514, 202

\bibitem[\protect\citeauthoryear{{Basko} \& {Sunyaev}}{{Basko} \&
  {Sunyaev}}{1976}]{BS1976}
{Basko} M.~M.,  {Sunyaev} R.~A.,  1976, \mnras, 175, 395

\bibitem[\protect\citeauthoryear{{Beaudet}, {Petrosian} \&
  {Salpeter}}{{Beaudet} et~al.}{1967}]{1967ApJ...150..979B}
{Beaudet} G.,  {Petrosian} V.,    {Salpeter} E.~E.,  1967, \apj, 150, 979

\bibitem[\protect\citeauthoryear{{Beloborodov}}{{Beloborodov}}{1998}]{1998MNRAS.297..739B}
{Beloborodov} A.~M.,  1998, \mnras, 297, 739

\bibitem[\protect\citeauthoryear{{Bildsten} \& {Brown}}{{Bildsten} \&
  {Brown}}{1997}]{1997ApJ...477..897B}
{Bildsten} L.,  {Brown} E.~F.,  1997, \apj, 477, 897

\bibitem[\protect\citeauthoryear{{Chashkina}, {Abolmasov} \&
  {Poutanen}}{{Chashkina} et~al.}{2017}]{2017arXiv1703.07005C}
{Chashkina} A.,  {Abolmasov} P.,    {Poutanen} J.,  2017, \mnras, 470, 2799

\bibitem[\protect\citeauthoryear{{Colbert} \& {Mushotzky}}{{Colbert} \&
  {Mushotzky}}{1999}]{1999ApJ...519...89C}
{Colbert} E.~J.~M.,  {Mushotzky} R.~F.,  1999, \apj, 519, 89

\bibitem[\protect\citeauthoryear{{Dall'Osso}, {Perna} \& {Stella}}{{Dall'Osso}
  et~al.}{2015}]{2015MNRAS.449.2144D}
{Dall'Osso} S.,  {Perna} R.,    {Stella} L.,  2015, \mnras, 449, 2144

\bibitem[\protect\citeauthoryear{{Ek{\c s}i}, {Anda{\c c}}, {{\c
  C}{\i}k{\i}nto{\u g}lu}, {Gen{\c c}ali}, {G{\"u}ng{\"o}r} \&
  {{\"O}ztekin}}{{Ek{\c s}i} et~al.}{2015}]{2015MNRAS.448L..40E}
{Ek{\c s}i} K.~Y.,  {Anda{\c c}} {\.I}.~C.,  {{\c C}{\i}k{\i}nto{\u g}lu} S.,
  {Gen{\c c}ali} A.~A.,  {G{\"u}ng{\"o}r} C.,    {{\"O}ztekin} F.,  2015,
  \mnras, 448, L40

\bibitem[\protect\citeauthoryear{{F{\"u}rst}
  et~al.}{2016}]{2016ApJ...831L..14F}
{F{\"u}rst} F. et al.,  2016, \apjl, 831, L14

\bibitem[\protect\citeauthoryear{{F{\"u}rst} et al.}{{F{\"u}rst}
  et~al.}{2017}]{2017ApJ...834...77F}
{F{\"u}rst} F. et al.,  2017, \apj, 834, 77

\bibitem[\protect\citeauthoryear{{Harding} \& {Lai}}{{Harding} \&
  {Lai}}{2006}]{2006RPPh...69.2631H}
{Harding} A.~K.,  {Lai} D.,  2006, Reports on Progress in Physics, 69, 2631

\bibitem[\protect\citeauthoryear{{Herold}}{{Herold}}{1979}]{1979PhRvD..19.2868H}
{Herold} H.,  1979, \prd, 19, 2868

\bibitem[\protect\citeauthoryear{{Illarionov} \& {Sunyaev}}{{Illarionov} \&
  {Sunyaev}}{1975}]{1975A&A....39..185I}
{Illarionov} A.~F.,  {Sunyaev} R.~A.,  1975, \aap, 39, 185

\bibitem[\protect\citeauthoryear{{Israel}}{{Israel et al.}}{2017a}]{Israel17Sci}
{Israel} G.~L. et al.,  2017a, Science, 355, 817

\bibitem[\protect\citeauthoryear{{Israel} et~al.}{2017b}]{Israel17MN}
{Israel} G.~L. et al.,  2017b, \mnras, 466, L48

\bibitem[\protect\citeauthoryear{{Kaminker}, {Gnedin}, {Yakovlev},
  {Amsterdamski} \& {Haensel}}{{Kaminker} et~al.}{1992}]{1992PhRvD..46.4133K}
{Kaminker} A.~D.,  {Gnedin} O.~Y.,  {Yakovlev} D.~G.,  {Amsterdamski} P.,
  {Haensel} P.,  1992, Phys.Rev.D, 46, 4133

\bibitem[\protect\citeauthoryear{{Kaminker}, {Gnedin}, {Yakovlev},
  {Amsterdamski} \& {Haensel}}{{Kaminker} et~al.}{1994}]{1994A&AT....4..283K}
{Kaminker} A.~D.,  {Gnedin} O.~Y.,  {Yakovlev} D.~G.,  {Amsterdamski} P.,
  {Haensel} P.,  1994, Astronomical and Astrophysical Transactions, 4, 283

\bibitem[\protect\citeauthoryear{{King}, {Lasota} \& {Klu{\'z}niak}}{{King}
  et~al.}{2017}]{2017MNRAS.468L..59K}
{King} A.,  {Lasota} J.-P.,    {Klu{\'z}niak} W.,  2017, \mnras, 468, L59

\bibitem[\protect\citeauthoryear{{Lipunov}}{{Lipunov}}{1982}]{1982SvA....26...54L}
{Lipunov} V.~M.,  1982, Soviet Astronomy, 26, 54

\bibitem[\protect\citeauthoryear{{Lyubarskii} \& {Syunyaev}}{{Lyubarskii} \&
  {Syunyaev}}{1988}]{LS88}
{Lyubarskii} Y.~E.,  {Syunyaev} R.~A.,  1988, Soviet Astronomy Letters, 14, 390

\bibitem[\protect\citeauthoryear{{Mineo}, {Gilfanov} \& {Sunyaev}}{{Mineo}
  et~al.}{2012}]{2012MNRAS.419.2095M}
{Mineo} S.,  {Gilfanov} M.,    {Sunyaev} R.,  2012, \mnras, 419, 2095

\bibitem[\protect\citeauthoryear{{Misiaszek}, {Odrzywo{\l}ek} \&
  {Kutschera}}{{Misiaszek} et~al.}{2006}]{2006PhRvD..74d3006M}
{Misiaszek} M.,  {Odrzywo{\l}ek} A.,    {Kutschera} M.,  2006, Phys.Rev.D, 74,
  043006

\bibitem[\protect\citeauthoryear{{Mushtukov}, {Nagirner} \&
  {Poutanen}}{{Mushtukov} et~al.}{2016}]{2016PhRvD..93j5003M}
{Mushtukov} A.~A.,  {Nagirner} D.~I.,    {Poutanen} J.,  2016, Phys.Rev.D, 93,
  105003

\bibitem[\protect\citeauthoryear{{Mushtukov}, {Suleimanov}, {Tsygankov} \&
  {Ingram}}{{Mushtukov} et~al.}{2017}]{2017MNRAS.467.1202M}
{Mushtukov} A.~A.,  {Suleimanov} V.~F.,  {Tsygankov} S.~S.,    {Ingram} A.,
  2017, \mnras, 467, 1202

\bibitem[\protect\citeauthoryear{{Mushtukov}, {Suleimanov}, {Tsygankov} \&
  {Poutanen}}{{Mushtukov} et~al.}{2015a}]{2015MNRAS.454.2539M}
{Mushtukov} A.~A.,  {Suleimanov} V.~F.,  {Tsygankov} S.~S.,    {Poutanen} J.,
  2015a, \mnras, 454, 2539

\bibitem[\protect\citeauthoryear{{Mushtukov}, {Suleimanov}, {Tsygankov} \&
  {Poutanen}}{{Mushtukov} et~al.}{2015b}]{2015MNRAS.447.1847M}
{Mushtukov} A.~A.,  {Suleimanov} V.~F.,  {Tsygankov} S.~S.,    {Poutanen} J.,
  2015b, \mnras, 447, 1847

\bibitem[\protect\citeauthoryear{{Parfrey}, {Spitkovsky} \&
  {Beloborodov}}{{Parfrey} et~al.}{2016}]{2016ApJ...822...33P}
{Parfrey} K.,  {Spitkovsky} A.,    {Beloborodov} A.~M.,  2016, \apj, 822, 33

\bibitem[\protect\citeauthoryear{{Poutanen}, {Fabrika}, {Valeev}, {Sholukhova}
  \& {Greiner}}{{Poutanen} et~al.}{2013}]{2013MNRAS.432..506P}
{Poutanen} J.,  {Fabrika} S.,  {Valeev} A.~F.,  {Sholukhova} O.,    {Greiner}
  J.,  2013, \mnras, 432, 506

\bibitem[\protect\citeauthoryear{{Poutanen}, {Lipunova}, {Fabrika}, {Butkevich}
  \& {Abolmasov}}{{Poutanen} et~al.}{2007}]{2007MNRAS.377.1187P}
{Poutanen} J.,  {Lipunova} G.,  {Fabrika} S.,  {Butkevich} A.~G.,
  {Abolmasov} P.,  2007, \mnras, 377, 1187
  
\bibitem[\protect\citeauthoryear{{Rumyantsev} \& {Chistyakov}}{{Rumyantsev} \&
  {Chistyakov}}{2008}]{2008JETP..107..533R}
{Rumyantsev} D.~A.,  {Chistyakov} M.~V.,  2008, Journal of Experimental and
  Theoretical Physics, 107, 533  

\bibitem[\protect\citeauthoryear{{Tsygankov}, {Lutovinov}, {Doroshenko},
  {Mushtukov}, {Suleimanov} \& {Poutanen}}{{Tsygankov}
  et~al.}{2016}]{2016A&A...593A..16T}
{Tsygankov} S.~S.,  {Lutovinov} A.~A.,  {Doroshenko} V.,  {Mushtukov} A.~A.,
  {Suleimanov} V.,    {Poutanen} J.,  2016, \aap, 593, A16

\bibitem[\protect\citeauthoryear{{Tsygankov}, {Mushtukov}, {Suleimanov} \&
  {Poutanen}}{{Tsygankov} et~al.}{2016}]{2016MNRAS.457.1101T}
{Tsygankov} S.~S.,  {Mushtukov} A.~A.,  {Suleimanov} V.~F.,    {Poutanen} J.,
  2016, \mnras, 457, 1101

\bibitem[\protect\citeauthoryear{{Wang} \& {Frank}}{{Wang} \&
  {Frank}}{1981}]{1981A&A....93..255W}
{Wang} Y.-M.,  {Frank} J.,  1981, \aap, 93, 255

\bibitem[\protect\citeauthoryear{{Yakovlev}}{{Yakovlev}}{1984}]{1984Ap&SS..98...37Y}
{Yakovlev} D.~G.,  1984, \apss, 98, 37

\bibitem[\protect\citeauthoryear{{Zeldovich} \& {Novikov}}{{Zeldovich} \&
  {Novikov}}{1975}]{1975seu..book.....Z}
{Zeldovich} I.~B.,  {Novikov} I.~D.,  1975, {Structure and evolution of the
  universe, Nauka, Moscow}

\end{thebibliography}

{

}

\appendix

\section{Neutrino energy losses in strong magnetic field}
\label{App:Nu}

\begin{table}
\centering   
\caption{Coefficients in equation (\ref{eq:Kaminker}). 
\label{tab1} 
}
\begin{tabular}{l c c c c c }
\hline
 $i:$ & 1 & 2 & 3 & 4 & 5 \\
 \hline\hline  
$a_i$ & 3.581  & 39.64   & 24.43   & 36.49  & 18.75 \\
$b_i$ & 1.058  & 0.6701   & 0.9143   & 0.472  & - \\
$c_i$ & $3.106\times 10^{-6}$  & $1.491\times 10^{-3}$   & $4.839\times 10^{-6}$   & -  & - \\
 \hline 
\end{tabular}
\end{table}

We used the approximation designed for the case of highly magnetized plasma obtained by \cite{1992PhRvD..46.4133K,1994A&AT....4..283K}, where the neutrino emission was calculated for the case of arbitrary magnetic field strength with the only assumption of nondegenerate plasma. 

Strong magnetic field modifies the conditions of plasma degeneracy because the magnetic field forces electrons to occupy ground or the lowest Landau levels (see \citealt{1984Ap&SS..98...37Y}). It results in essentially reduced Fermi momentum
$$p_{\rm F}=\frac{h}{2}\frac{hc}{eB}n_{\rm e}\simeq 1.37\times 10^{-45}n_{\rm e}B_{12}^{-1}\,\,{\rm erg\,s\,cm^{-1}}$$
as compared to the field-free case $p_{\rm F}=h(3n_{\rm e}/8\pi)^{1/3}\simeq 3.26\times 10^{-27}n_{\rm e}^{1/3}\,\,{\rm erg\,s\,cm^{-1}}$ and reduced Fermi temperature $T_{\rm F}$. 

The main condition of the approximation which we use is sufficiently high temperature: $T\gg T_{\rm F}$. The Fermi temperature as a function of mass density was calculated by many authors (see Fig.10 in review by \citealt{2006RPPh...69.2631H}). The mass density $\rho<10^{5}\,{\rm g\, cm^{-3}}$ (which is reasonable upper limit for the mass density in accretion column) the expected temperatures $\sim 10^{10}\,{\rm K}$ are well above actual Fermi temperature $T_{\rm F}$. Therefore, the calculations performed by  \cite{1992PhRvD..46.4133K} are valid and we can use approximation:
\beq\label{eq:Kaminker}
Q^{-}_{\nu}=\frac{Q_{\rm c}}{\pi^4}
\left\{\left[t^3\overline{C^2_V}(1+3.75t)+t^4(\overline{C^2_V}+\overline{C^2_A})P(t)\right]F(t,b )\right. \nonumber \\
\left. +
\frac{tb^2}{6(1+b)}\left[\overline{C^2_V}+\overline{C^2_A}\frac{b}{1+b}\right]S(t)
\right\}
\exp\left[-\frac{2}{t}\right],
\eeq
where
$Q_{\rm c}=1.015\times 10^{23}\,\,{\rm erg\,cm^{-3}\,s^{-1}}$ and dimensionless $B$-field strength $b$
and temperature $t$ are given by
$$b=\frac{B}{4.414\times 10^{13}\,{\rm G}},\quad\quad t=\frac{T}{5.93\times 10^{9}\,{\rm K}}$$
and
$$P(t)=1+\sum\limits_{i=1}^{5}a_i t^i,\quad\quad S(t)=1+\sum\limits_{i=1}^{4}b_i t^i,$$
$$F(t,b)=\frac{1}{R_1 R_2 R_3},\quad\quad R_i=1+c_i\frac{b}{t^2}\exp\left[\frac{(2b)^{1/2}}{3t}\right]$$
with approximate coefficients presented in Table \ref{tab1}. If the temperature $10^9\,{\rm K} < T < 10^{11}\,{\rm K}$ and magnetic field strength $ B <10^{15}\,{\rm G}$ the approximation reduces to rough equation (\ref{eq:Qnu}) given in our paper (see also \citealt{1994A&AT....4..283K}).

It was shown by \cite{1992PhRvD..46.4133K} that the neutrino emissivity due to the pair-annihilation process in nondegenerate plasma is independent on density. It seems that it is the only density independent process of neutrino creation throughout the non-degenerate plasma.

\bsp	
\label{lastpage}
\end{document}